# U-Shaped Fermi-Level Dependence of Point-Defect Aggregation Enthalpy in Silicon

*Yin Wu*[1], *Xiao Kong*[*1], *Pai Li*[*1]

[1]*Shanghai Institute of Microsystem and Information Technology, Chinese Academy of Sciences Shanghai 200050, China*

***Corresponding Author**

Email: xkong91@mail.sim.ac.cn; lipai@mail.sim.ac.cn

**ABSTRACT**

Point defects govern the electronic and structural behavior of silicon, yet their configuration and energy depend on both the charge state and the Fermi level ($E_F$). Using an unbiased global structure search with first-principles calculations, we screened more than 1,500 candidate structures, covering intrinsic defects and C, H, O, N, P, and B impurities in charge states from −2 to +2. The reaction enthalpy of defect aggregation depends on $E_F$ in a U shape, being strongest where the net charge transferred during the reaction vanishes and weaker toward both band edges. This U arises because complexes are more charge-neutral than their isolated constituents, so the net charge transfer changes sign across the gap. The point where this sign change occurs is set by the transition levels of the specific defects. For typical reactions, the driving force is tunable by 0.3–0.5 eV, which shifts the equilibrium complex concentration by five to eight orders of magnitude at room temperature. Our results make the Fermi level a practical lever for defect engineering in as-grown and irradiated silicon.

## INTRODUCTION

Point defects govern the electrical behavior of silicon. They set doping, mediate carrier recombination, and shape the microdefect distribution, so they decide device design and reliability[1–3] and device performance.[4,5] A defect, however, is not a fixed object. Its stable structure and its formation energy both depend on its charge state, and the charge state is set by the Fermi level. This coupling is the defining complication, because it means that moving the Fermi level, which doping or an applied field does, can change a defect's structure and its energy. Whether the Fermi level can truly be used to control defects, however, has remained an open question.

Configuration is the link between charge state and behavior. Because the geometry sets both the electronic levels and the formation energy, a defect whose structure changes with charge state also changes its stability and its electrical activity. Within this framework, silicon is, in particular, the fundamental substrate for integrated circuits and photovoltaics.[6,7] Even in this simple material, experiments have long observed the defect multi-stability. The self-interstitial adopts three distinct structures,[8] a phosphorus-related defect in silicon shows four charge states,[9,10] and carbon- and hydrogen-related centers change configuration with their charge state.[11–14] The coupling is therefore both real and widespread.

Of the processes this coupling affects, aggregation is the most consequential. Isolated defects bind into complexes, and complexes drive microdefect growth and dopant deactivation. Whether a complex forms is set by the reaction enthalpy between the isolated defects and the complex. Because this enthalpy also depends on the Fermi level, the Fermi level becomes a direct lever on aggregation and on the defect concentration.[15–21]

Experiments cannot resolve this coupling. Electron paramagnetic resonance,[22] deep-level transient spectroscopy,[23] and photoluminescence[24] give only indirect information. They lose the atomic structure when several configurations coexist or a charge-state change rearranges the defect. Theory has not closed the gap. Most models start from a few hand-built structures,[25] so they cannot survey the charge-state-dependent potential energy surface. As a result, there is no rule for how the Fermi level sets the aggregation driving force.

Here we combine an unbiased global structure search[26] with first-principles calculations to map point defects and their complexes in silicon. We cover C, H, O, N, P, and B impurities over charge states from −2 to +2. We find that the aggregation reaction enthalpy follows a U-shaped Fermi-level dependence, so the Fermi level is a quantitative lever on whether complexes form. We further classify defects by their configurational response and identify negative-U centers, and we show that aggregation can shift the Fermi level in return.

## RESULTS AND DISCUSSION

### 1. Unbiased global structure search of point defects in silicon

The workflow in **Figure 1** covers intrinsic defects (vacancies and self-interstitials), the common impurities C, H, O, N, B, and P, and their complexes. We chose these for their relevance to silicon devices and their diverse atomic behavior. The global structure search uses the stochastic surface walking (SSW) method[27] at the PBE level.[28] It does not prescribe a starting geometry; instead, it samples many candidate structures, which can uncover geometries that a hand-built model would not include. The search runs at the neutral charge state and screens more than 1,500 candidate structures (**Figure S1**). For each defect, the 10 most stable configurations are then re-optimized with the higher-accuracy SCAN functional[29] and the vdW-D3(BJ) correction,[30] across the five charge states from −2 to +2.

We use a naming convention based on defect type. An interstitial atom is written $I_X$, with X the element symbol, and two aggregated interstitials are $(I_X)_2$. Substitutional impurities are $X_{Si}$. Vacancies are written V, and n vacancies are $V_n$. Composite defects appear in brackets, for example $[I_B \backslash (I_O)_2 \backslash V_2]$, which gathers one boron interstitial, two oxygen interstitials, and two vacancies.

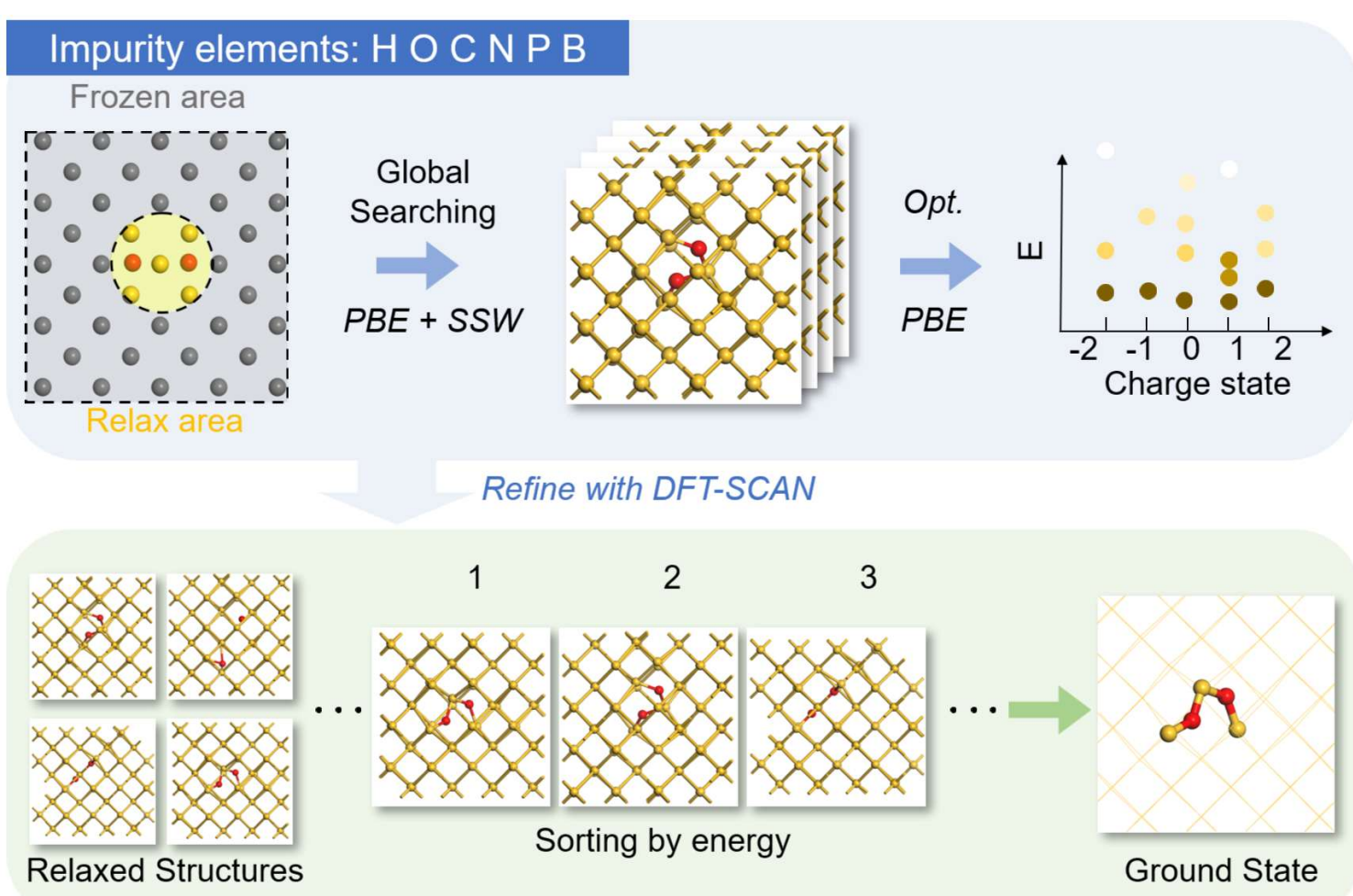

**Figure 1.** Overview of the workflow for exploring and characterizing point defects in silicon.

## 2. Charge-state-dependent configuration and electronic properties of point defects

The ground-state structures of each defect at different charge states appear in **Figure S2**. They fall into two classes, reconfigurable and non-reconfigurable, based on how much the charge state distorts the structure. Their formation energies are plotted in **Figure S3**. Among the isolated defects, carbon-related species generally have lower formation energies than nitrogen species or silicon vacancies. When these species form complexes such as [$I_C$\($I_O$)$_2$\$V_2$] or [$I_B$($I_O$)$_2$\$V_2$], the formation energies rise, so these complexes are less stable than the isolated species.

The charge-state transition levels are given by the crossing points of the formation-energy curves of the stable charge configurations (**Figure 2**a). For $C_{Si}$, $I_O$ and [$I_C$\$I_O$\$V_{Si}$], the most stable charge state does not change across the gap, so there is no transition level and the defect holds a fixed charge. The other defects have transition levels inside the gap and fall into two groups. In a continuous transition, the stable charge state changes one step at a time as the Fermi level moves, skipping no charge state. Typical examples are $I_C$, $H_{Si}$ and $V_{Si}$. In a jumping transition, some charge states are never stable, so the level order jumps. This jump is a signature of negative-U behavior.[31] It appears in the $\varepsilon(+1/-1)$ level of $I_H$ and $I_B$, and the $\varepsilon(+2/0)$ level of $(I_N)_2$.

The charge state also shifts the ground-state configuration of a defect,[32] and the size of this change varies among defects. We quantify the configurational diversity by the deviations in bond length and bond angle between charge states (**Figure 2**b). Two groups emerge. Non-reconfigurable defects show small bond-length and bond-angle changes at all Fermi levels, with no bond rearrangement. The $[I_C \backslash C_{Si}]$ defect is a clear example. Its C-Si bond lengths stay between 1.84 and 1.94 Å, and its bond angles stay near 131° to 133°, across charge states from −2 to +2 (**Figure S4**). Reconfigurable defects show strong configurational sensitivity. In $[I_B \backslash (I_O)_2 \backslash V_2]$, the O-O bond length grows from 1.52 to 1.78 Å as the charge state goes from +1 to −2. The boron coordination changes from $sp^3$ to $sp^2$ (**Figure S5**). All defects with negative-U behavior belong to the reconfigurable class and show large configurational diversity, as highlighted in **Figure 2**b.

The two classes differ mainly in the strength of electron-phonon deformation potential coupling[33] . For a weakly coupled defect such as the carbon interstitial, adding or removing an electron causes only a small structural change, with no change in coordination (**Figure S6**). The neutral defect is a closed shell with zero magnetic moment. Removing one electron lengthens all C-Si bonds uniformly through a Jahn-Teller distortion,[24] which preserves the bond topology.

As a general trend, deep-level defects behave as donors when $E_F$ is near the VBM, donating electrons, and as acceptors when $E_F$ is near the CBM, capturing electrons. The average charge is therefore positive at the VBM and negative at the CBM (**Figure 2**c).

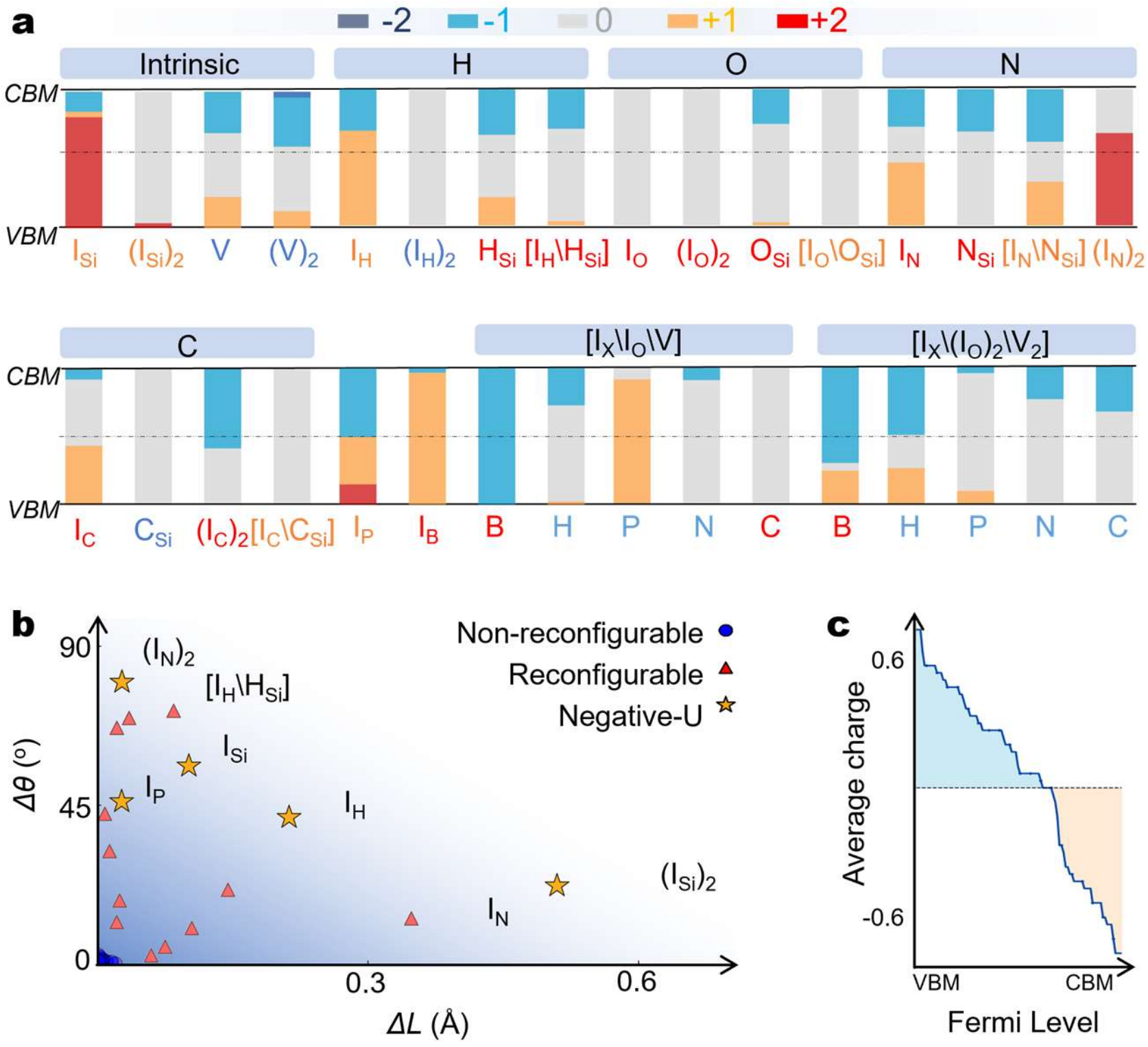

**Figure 2.** Configuration-diversity analysis of point defects. (a) Charge-state transition levels for selected point defects across the gap between the VBM and CBM, spanning charge states from −2 to +2. For the sets $[I_X\backslash I_O\backslash V]$ and $[I_X\backslash(I_O)_2\backslash V_2]$, X is B, H, P, N, or C as shown. (b) Configurational diversity at different charge states, from bond-length and bond-angle deviations. Defects with negative-U behavior are marked with stars. (c) Average charge of all the studied point defects as a function of the Fermi level.

## 3. U-shaped Fermi-level dependence of the aggregation reaction enthalpy

Aggregation converts two isolated defects into a complex. We describe the thermodynamic driving force with the reaction enthalpy of A + B → [A\B]. It is the formation energy of the complex minus the formation energies of the two reactants:

$$\Delta \mathrm{H}(E_F) = E^f([A\backslash B], E_F) - E^f(A, E_F) - E^f(B, E_F)$$

where $E^f(X, E_F)$ is the formation energy of defect $X$ at Fermi level $E_F$ referenced to the VBM. The reaction also exchanges charge with the host (**Figure S8**). The net charge transfer of the reaction is the charge state of the complex minus the sum of the reactant charge states:

$$\Delta Q(E_F) = q([A\backslash B], E_F) - q(A, E_F) - q(B, E_F)$$

where $q(X, E_F)$ is the charge state of defect $X$ at $E_F$.

Aggregation is generally favorable across the gap, as the negative reaction enthalpies in **Figure 3**a indicate. The origin is a systematic reduction in charge, since binding gathers the reactants into a more charge-neutral complex (**Figure 3**b). Near the VBM, the complex tends to be less positive than its reactants, and near the CBM it tends to be less negative. The net charge transfer is therefore negative near the VBM and positive near the CBM (**Figure 3**c).

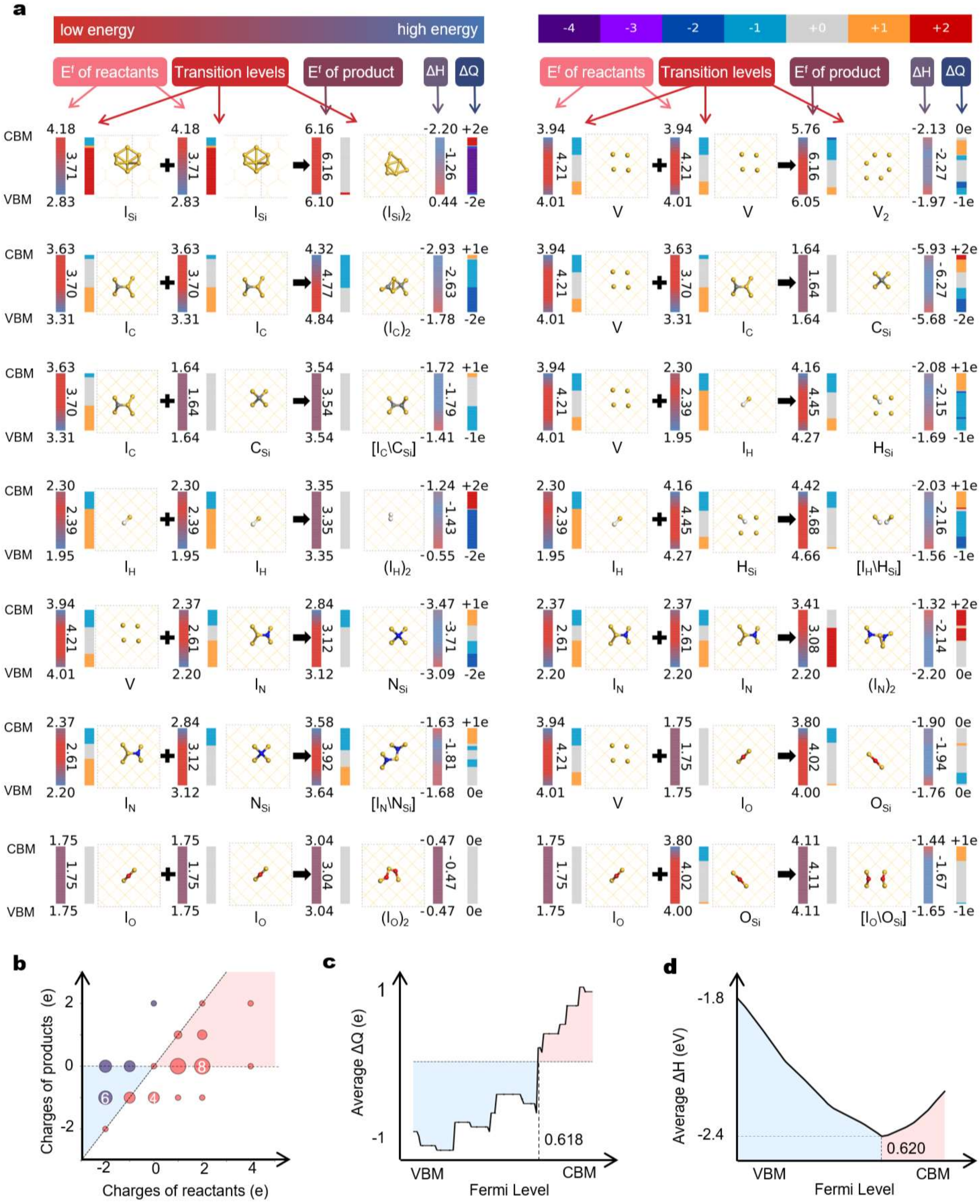

**Figure 3.** Thermodynamics of aggregation. (a) For each reaction, the structures and the Fermi-level-dependent formation energies of the two isolated defects and of the complex, together with the reaction enthalpy and the net charge transfer. (b) Distribution of the reactions in the charge-state space of reactants and products; the bubble diameter is proportional to the number of reactions at that combination. (c) Average net charge transfer and (d) average reaction enthalpy as a function of the Fermi level. Each point is the mean over the fourteen reactions at that level.

The reaction enthalpy is piecewise linear in $E_F$, with the intervals set by the charge-state transition levels. Because each formation energy depends on $E_F$ with a slope equal to its charge state, and the charge state is constant within an interval, the slope of ΔH equals ΔQ there (**Figure 4**a). Within each interval, ΔH therefore takes the form:

$$\Delta \mathrm{H} = \Delta \mathrm{H}^{const} + \Delta Q \times E_F$$

where $\Delta H^{const}$ gathers the $E_F$-independent contributions of that segment. A negative slope near the VBM and a positive slope near the CBM therefore force a minimum where ΔQ changes sign. The enthalpy therefore falls from the VBM toward the charge-neutral point, then rises toward the CBM. Across the studied reactions the average traces a U shape across the gap (**Figure 3**d).

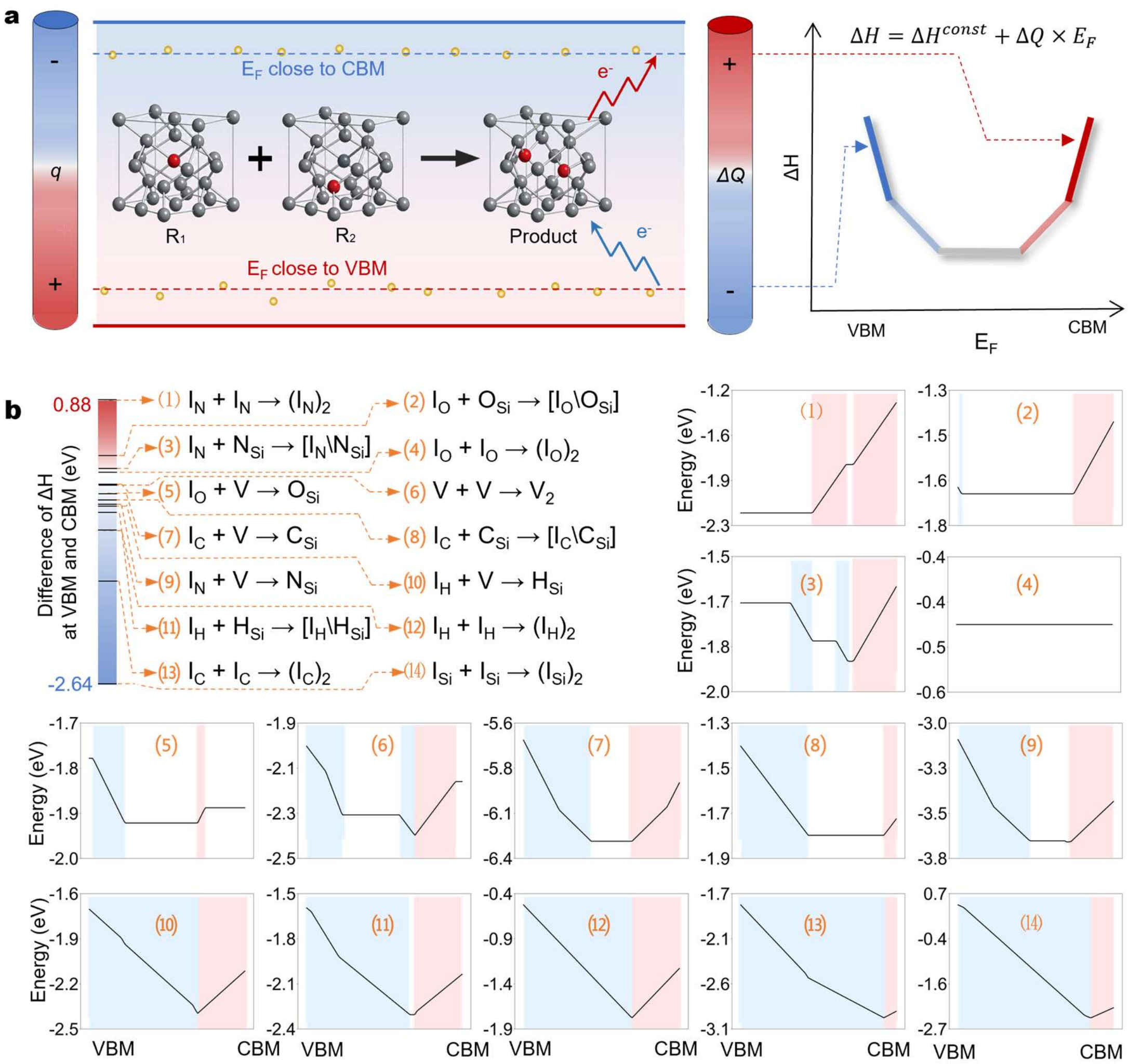

**Figure 4.** Mechanism and Fermi-level dependence of point-defect aggregation. (a) Schematic of the mechanism behind the U-shaped reaction enthalpy. The enthalpy takes the form $\Delta H = \Delta H^{const} + \Delta Q \times E_F$, so its slope is the net charge transfer $\Delta Q$. (b) Reaction enthalpy versus $E_F$ for each of the fourteen reactions, ordered by the difference between the band-edge values, $\Delta H$ at the CBM minus ΔH at the VBM.

The individual reactions do not all take this shape (**Figure 4**b). Of the fourteen reactions, three show a clear interior minimum, namely $I_N + N_{Si}$, V + V, and $I_C$ + V. The rest are monotonic, with a single extremum at one band edge. The U shape of **Figure 3**d is an average over this ensemble. Whether a single reaction is U shaped or L shaped is set by its charge-state ordering. A true U requires the net charge transfer to change sign inside the gap, which happens when the transition levels of the complex lie between those of the two reactants. When the complex keeps a one-signed charge offset across the whole gap, the curve stays L shaped.

The U shape here is not perfect symmetric. For several reactions, such as $I_C + I_C \rightarrow (I_C)_2$ and $I_{Si} + I_{Si} \rightarrow (I_{Si})_2$, the two arms differ in magnitude. The difference between the band-edge values, which is the value at the CBM minus the value at the VBM, is shown by the color scale of **Figure 4**b. This asymmetry reflects the work function of silicon and the asymmetric placement of the transition levels in the gap. As a result, the minimum of a given U does not sit at the geometric mid-gap. In the present set the average charge-neutral point, where the net charge transfer averages to zero, lies somewhat toward the CBM.

The size of the effect sets its practical value. The reaction enthalpy changes by roughly 0.3 to 0.5 eV for typical reactions between its maximum and minimum. The equilibrium concentration of a complex scales exponentially with the negative of its formation energy divided by thermal energy, so it is largest where ΔH is most negative and smallest where $\Delta H$ is least negative. The tunability of the complex concentration is the ratio of these two extreme values: $\exp[(\Delta H^{max} - \Delta H^{min})/kT]$, where $\Delta H^{max}$ and $\Delta H^{min}$ are the maximum and minimum of $\Delta H$ in the band gap, $k$ is the Boltzmann constant, and $T$ is the temperature. The chemical potential and other constant terms cancel in this ratio. A change of 0.3 to 0.5 eV therefore shifts the concentration by about five to eight orders of magnitude at room temperature. This quantity is shown in **Figure 5**. The largest tunability appears for $(I_{Si})_2$, $(I_H)_2$, and $(I_C)_2$, which respond most strongly near the conduction band. The Fermi level, or equivalently the n-type or p-type doping level, is thus a quantitative lever for the abundance of aggregates.

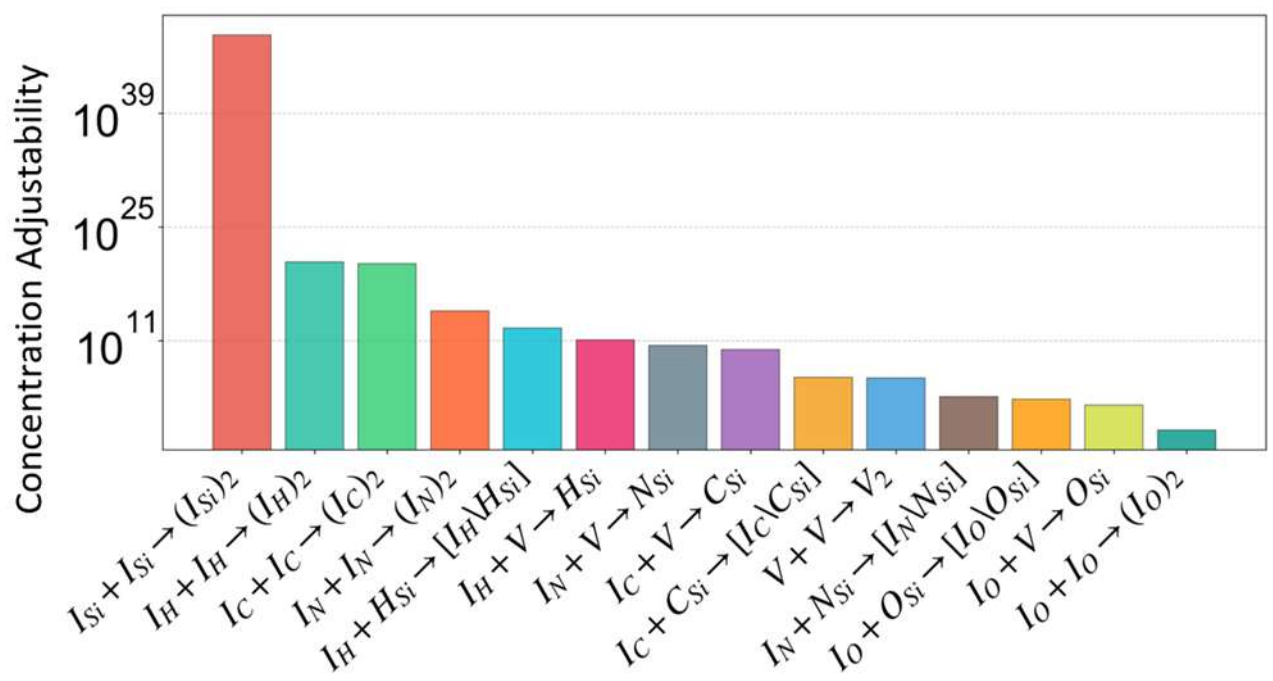

**Figure 5.** Tunability of the equilibrium concentration of each complex by Fermi-level control.

Two modeling choices bound these results. First, the formation energies are zero-temperature enthalpies. Vibrational and configurational entropy are not included. Second, the values are computed at fixed chemical potentials. A different chemical potential shifts the baseline enthalpy but does not change the U-shaped slope structure. Third, the SCAN gap of 0.89 eV is smaller than the experimental gap of 1.12 eV. The absolute positions of the transition levels therefore carry a systematic offset, while the relative trends are unaffected.

Finally, aggregation is not a passive response to a preset Fermi level. When a complex forms, the net charge transfer exchanges charge with the host. A negative transfer at the VBM side removes electrons from the host and raises $E_F$. Because the enthalpy falls as $E_F$ rises toward the minimum, a higher $E_F$ strengthens the driving force for further aggregation. The same positive feedback acts on the CBM side, where a positive transfer releases electrons, lowers $E_F$, and again moves the system toward the minimum. This self-consistent loop matters only when the defect concentration is large enough to shift $E_F$ relative to the shallow dopants. A well-known case is oxygen precipitation and thermal donor formation in Czochralski silicon,[34] which couple aggregation to the Fermi level. Charge transfer therefore lets aggregation engineer its own electrostatic environment in a self-driven way.

The mechanism rests only on charge neutralization, a generic electrostatic tendency, so it should extend to other semiconductors that host charge-state-multistable defects. In wide-gap materials, where native defects often set the reachable carrier type,[35,36] this coupling of aggregation to the Fermi level may prove even more consequential than in silicon.

## CONCLUSION

In summary, we combined an unbiased global structure search with first-principles calculations to map point defects and their complexes in silicon. The map spans charge states from −2 to +2 for C, H, O, N, P, and B impurities. It reveals a U-shaped Fermi-level dependence of the aggregation reaction enthalpy. Aggregation is strongest where the net charge transferred vanishes and weaker toward the band edges, since complexes are more charge-neutral than their isolated defects. This driving force, tunable by 0.3 to 0.5 eV, shifts the equilibrium complex concentration by five to eight orders of magnitude at room temperature. The Fermi level is therefore a quantitative lever for microdefect engineering in silicon, and aggregation can shift the Fermi level in return by exchanging charge with the host. We expect the mechanism to extend to other semiconductors with charge-state-multistable defects, although the exact U shape is set by the charge-state ordering in each material.

## METHODS

**DFT calculation.** All DFT calculations were carried out using the Vienna *ab initio* simulation package (VASP) version 6.4.3.[37,38] For global search, the PBE functional is adopted. The kinetic energy cutoff is set to 450 eV for systems without oxygen and 540 eV for systems containing oxygen. For further optimization of the 10 most stable candidate defect structures in each case, the strongly constrained and appropriately normed (SCAN) functional[29] is adopted with a vdw-D3 (BJ) correction.[39,40] The projector augmented wave method[41] is used to model core-valence interactions. A kinetic energy cutoff of 540 eV is used for the plane-wave basis set for all SCAN calculations. The energy convergence threshold is set to $1\times10^{-5}$ eV, and atomic positions are relaxed using the conjugate gradient method until the force on each atom was below $10^{-2}$ eV/Å. The reciprocal space is sampled using a Γ-centered Monkhorst-Pack K-mesh with grid spacings smaller than $0.02\times2\pi$ Å$^{-1}$. Each defect is placed at the center of a silicon supercell, which is divided into two regions. A mobile core region holds the defect atoms, and these atoms move throughout the calculation. A frozen outer region keeps the surrounding host atoms fixed during the SSW steps and releases them for the structural optimization, so that every atom is free to relax at that stage. In all steps the cell shape and volume are kept at the bulk values. This prevents the defect from interacting elastically with its periodic images.[42] Spin polarization is explicitly included in all calculations, as previous studies have demonstrated its critical importance—neglecting spin polarization can lead to errors of up to a factor of two in defect energies, as exemplified by the hydrogen defect in silicon.[43] Each defect is placed at the center of a silicon supercell split into two regions: a mobile core region containing the defect atoms, where all atoms are allowed to move during the global search, and a frozen outer region where atoms were fixed during SSW steps and movable at the structural optimization steps.

**Global search for point defect structures.** In this work, all global structural searches were performed with the stochastic surface walking (SSW) method[27] at the PBE level via the VASP interface. The SSW calculations are carried out using the following parameters: the number of SSW steps per walk was set to 15, the force convergence criterion is 0.1 eV/Å, and the root-mean-square displacement tolerance for structural convergence is 0.2 Å. The simulation temperature is fixed at 1000 K to promote crossing between metastable states during the search.

**Defect geometry analysis with respect to charge states.** To quantify the structural response of defects to changes in charge state (**Figure 2**b), we define two metrics relative to the neutral state. The normalized bond-length deviation $\Delta L$ is defined as:

$$\Delta L = \frac{1}{4}\sum_{q\in\{+1,+2,-1,-2\}}\frac{1}{N_b}\left|\sum_{i=1}^{N_b}L_i(q)-\sum_{i=1}^{N_b}L_i(0)\right|$$

where $L_i(q)$ is the length of the $i$-th bond in charge state $q$, and $N_b$ is the total number of bonds in the defect core. The bond-angle deviation is defined as the maximum over the four charge states:

$$\Delta\theta = \max_{q\in\{+1,+2,-1,-2\}} \frac{1}{N_a}\left|\sum_{i=1}^{N_a} \theta_i(q) - \sum_{i=1}^{N_a} \theta_i(0)\right|$$

where $\theta_i(q)$ is the $i$-th bond angle (in degrees) and $N_a$ is the number of bond angles. Based on these metrics, defects are classified as rigid if $\Delta L < 0.05$ Å and $\Delta\theta < 3°$, and as dynamic otherwise.

The different statistical treatments for bond lengths and bond angles reflect the distinct geometric characteristics of the defect core. While multiple bond lengths are present, the defect core typically contains only one characteristic bond angle (e.g., the Si-N-Si angle for nitrogen or the C-Si-C angle for carbon). For bond lengths, we use the average deviation across all four charge states to capture the overall structural response. For the single bond angle, we instead take the maximum deviation across charge states to emphasize the most extreme angular distortion, as large bending can occur in specific charge states (e.g., the Jahn–Teller distortion in $I_C$ or the coordination change in $I_N$) without being averaged out.

**Formation energy of charged defects.** The stability of a point defect in charge state $q$ is closely related to the Fermi level, and is evaluated as:

$$E_q^f(\boldsymbol{R_q}, E_F) = E_{\text{defect}}(\boldsymbol{R_q}, q) - E_{\text{host}} - \sum_i n_i\mu_i + qE_F + E_{\text{corr}}(\boldsymbol{R_q}, q)$$

where $\boldsymbol{R_q}$ is the ground-state structure of the defect $X$ in charge state $q$. $E_F$ is the Fermi level referenced to the VBM. $E_{\text{defect}}$ is the DFT energy of the defect system in charge state $q$, and $E_{\text{host}}$ is the DFT energy of the host crystalline silicon system under charge-neutral conditions. $n_i$ denotes the number of atoms added to ($n_i$>0) or removed from ($n_i$<0) the supercell to form the defect, and $\mu_i$ is the corresponding chemical potential. The charge state of the defect is denoted by $q$, and $E_{\text{corr}}$ is the correction term that accounts for electrostatic finite-size effects and potential alignment between the charged defect and bulk reference. The necessity for these corrections arises from a key artifact of the supercell approach: periodic images interact, so the supercell represents a finite defect concentration rather than the dilute limit. To correct for these artifacts, we employ the post-processing scheme proposed by Kumagai and Oba[44] for calculating $E_{\mathrm{corr}}$. The formation energy of a point defect $X$ is thus given by:

$$E^f(X, E_F) = \min_q E_q^f(\boldsymbol{R_q}, E_F)$$

and the charge state of defect $R$ at Fermi level $E_F$ is:

$$q(X, E_F) = arg \min_q E_q^f(\boldsymbol{R_q}, E_F)$$

**Chemical potentials in formation energy calculations.** The chemical potentials are referenced to their most stable elemental phases under ambient conditions: diamond for C, β-rhombohedral boron for B, and black phosphorus for P. For solid phases (Si, C, B, P), $\mu = E_{\mathrm{bulk}}/N$, where $E_{\mathrm{bulk}}$ is the DFT total energy of the conventional cell and $N$ is the number of atoms. For gaseous species ($N_2$, $H_2$), the chemical potential depends on temperature and pressure. The chemical potential of hydrogen is derived from the $H_2$ molecule as follows (similar for $N_2$):

$$\mu_{\mathrm{H}} = \frac{1}{2}\left[E_{H_2}^{DFT} + H_{\mathrm{H_2}}(T) - TS_{\mathrm{H_2}}(T) + k_{\mathrm{B}}T \ln\left(\frac{P_{\mathrm{H_2}}}{P_0}\right)\right]$$

where $E_{H_2}^{DFT}$ denotes the DFT energy of an isolated $H_2$ molecule. The terms $H_{H_2}(T)$ and $TS_{H_2}(T)$ represent the enthalpy and entropy of $H_2$ at temperature T, respectively, taken from standard thermodynamic data tables NIST-JANAF[45] to account for the temperature dependence of chemical potential. The final term $k_B T \, ln\left(\frac{P_{H_2}}{P_0}\right)$ represents the pressure correction, where $P_{H_2}$ is the partial pressure of hydrogen and $P_0$ is the standard pressure. For oxygen, $\mu_{\mathrm{O}}$ is derived from $SiO_2$, reflecting its origin from molten quartz crucible during silicon crystal growth. Specifically, the chemical potential of oxygen is given by:

$$\mu_{\mathrm{O}} = \frac{1}{2}\left(E_{SiO_2}^{DFT} - E_{Si}^{DFT}\right)$$

where $E_{SiO_2}^{DFT}$ and $E_{Si}^{DFT}$ are the DFT energy per formula unit of $\alpha$-quartz $SiO_2$ and crystalline Si, respectively.

**Charge-state transition level calculation.** The charge-state transition level, denoted as ε(q/q'), is the specific position of the Fermi level within the band gap at which the most stable charge of a defect from q to q'. It is calculated as:

$$\varepsilon(q/q') = \frac{E_{q'}^{f}\left(\boldsymbol{R}_{\boldsymbol{q}'}, E_F = 0\right) - E_{q}^{f}\left(\boldsymbol{R}_{\boldsymbol{q}}, E_F = 0\right)}{q - q'}$$

where $E_F = 0$ corresponds to the VBM level as reference. As discussed above, the most stable defect structures in different charge are not necessarily identical, and the expression can be equivalently decomposed as:

$$\varepsilon(q/q') = E_{\mathrm{QP}} + E_{\mathrm{relax}}$$

$$E_{\mathrm{QP}} = \frac{E_{q'}^{f}\left(\boldsymbol{R}_{\boldsymbol{q}'}, E_F = 0\right) - E_{q}^{f}\left(\boldsymbol{R}_{\boldsymbol{q}'}, E_F = 0\right)}{q - q'}$$

$$E_{\mathrm{relax}} \frac{E_{q}^{f}\left(\boldsymbol{R}_{\boldsymbol{q}'}, E_F = 0\right) - E_{q}^{f}\left(\boldsymbol{R}_{\boldsymbol{q}}, E_F = 0\right)}{q - q'}$$

where the first term $E_{\mathrm{QP}}$ is the energy cost to add an electron while keeping all atoms fixed, and it directly measures the electron–phonon coupling strength[46]. The second term $E_{\mathrm{relax}}$ accounts for lattice relaxation following electron addition, and is calculated as the energy difference between two configurations with the same charge state.

**Data availability**

The data supporting the findings of this study are available from the corresponding author upon reasonable request.

**REFERENCES**

(1) Sinno, T.; Dornberger, E.; Von Ammon, W.; Brown, R. A.; Dupret, F. Defect Engineering of Czochralski Single-Crystal Silicon. *Mater. Sci. Eng.: R: Rep.* **2000**, *28* (5–6), 149–198. https://doi.org/10.1016/S0927-796X(00)00015-2
(2) Meda, L.; Cerofolini, G. F.; Queirolo, G. Impurities and Defects in Silicon Single Crystal. *Prog. Cryst. Growth Charact.* **1987**, *15* (2), 97–134. https://doi.org/10.1016/0146-3535(87)90003-7
(3) Fahey, P. M.; Griffin, P. B.; Plummer, J. D. Point Defects and Dopant Diffusion in Silicon. *Rev. Mod. Phys.* **1989**, *61* (2), 289–384. https://doi.org/10.1103/RevModPhys.61.289
(4) Yang, Z.; Sakaguchi, N.; Watanabe, S.; Kawai, M. Dislocation Loop Formation and Growth under in Situ Laser and/or Electron Irradiation. *Sci. Rep.* **2011**, *1* (1), 190. https://doi.org/10.1038/srep00190
(5) Liu, Y.; Ding, Y.; Xie, J.; Xu, L.; Wha Jeong, I.; Yang, L. One-Step Femtosecond Laser Irradiation of Single-Crystal Silicon: Evolution of Micro-Nano Structures and Damage Investigation. *Mater. Des.* **2023**, *225*, 111443. https://doi.org/10.1016/j.matdes.2022.111443
(6) Fisher, G.; Seacrist, M. R.; Standley, R. W. Silicon Crystal Growth and Wafer Technologies. *Proc. IEEE* **2012**, *100* (Special Centennial Issue), 1454–1474. https://doi.org/10.1109/JPROC.2012.2189786
(7) Lu, J.; Kovalgin, A. Y.; Van Der Werf, K. H. M.; Schropp, R. E. I.; Schmitz, J. Integration of Solar Cells on Top of CMOS Chips Part I: A-Si Solar Cells. *IEEE Trans. Electron Devices* **2011**, *58* (7), 2014–2021. https://doi.org/10.1109/TED.2011.2143716
(8) Gharaibeh, M.; Estreicher, S. K.; Fedders, P. A. Molecular-Dynamics Studies of Self-Interstitial Aggregates in Si. *Physica B* **1999**, *273–274*, 532–534. https://doi.org/10.1016/S0921-4526(99)00566-9
(9) Chantre, A.; Kimerling, L. C. Configurationally Multistable Defect in Silicon. *Appl. Phys. Lett.* **1986**, *48* (15), 1000–1002. https://doi.org/10.1063/1.96669
(10) Zhan, X. D.; Watkins, G. D. Electron Paramagnetic Resonance of Multistable Interstitial-Carbon–Substitutional-Group-V-Atom Pairs in Silicon. *Phys. Rev. B* **1993**, *47* (11), 6363–6380. https://doi.org/10.1103/PhysRevB.47.6363
(11) Song, L. W.; Zhan, X. D.; Benson, B. W.; Watkins, G. D. Bistable Interstitial-Carbon–Substitutional-Carbon Pair in Silicon. *Phys. Rev. B* **1990**, *42* (9), 5765–5783. https://doi.org/10.1103/PhysRevB.42.5765
(12) Sah, C.-T.; Sun, J. Y.-C.; Tzou, J. J.-T. Deactivation of the Boron Acceptor in Silicon by Hydrogen. *Appl. Phys. Lett.* **1983**, *43* (2), 204–206. https://doi.org/10.1063/1.94287
(13) Johnson, N. M.; Herring, C.; Chadi, D. J. Interstitial Hydrogen and Neutralization of Shallow-Donor Impurities in Single-Crystal Silicon. *Phys. Rev. Lett.* **1986**, *56* (7), 769–772. https://doi.org/10.1103/PhysRevLett.56.769
(14) Van De Walle, C. G.; Denteneer, P. J. H.; Bar-Yam, Y.; Pantelides, S. T. Theory of Hydrogen Diffusion and Reactions in Crystalline Silicon. *Phys. Rev. B* **1989**, *39* (15), 10791–10808. https://doi.org/10.1103/PhysRevB.39.10791
(15) Freysoldt, C.; Grabowski, B.; Hickel, T.; Neugebauer, J.; Kresse, G.; Janotti, A.; Van De Walle, C. G. First-Principles Calculations for Point Defects in Solids. *Rev. Mod. Phys.* **2014**, *86* (1), 253–305. https://doi.org/10.1103/RevModPhys.86.253
(16) Broberg, D.; Bystrom, K.; Srivastava, S.; Dahliah, D.; Williamson, B. A. D.; Weston, L.; Scanlon, D. O.; Rignanese, G.-M.; Dwaraknath, S.; Varley, J.; Persson, K. A.; Asta, M.; Hautier, G. High-Throughput Calculations of Charged Point Defect Properties with Semi-Local Density Functional

Theory—Performance Benchmarks for Materials Screening Applications. *npj Comput Mater* **2023**, *9* (1), 72. https://doi.org/10.1038/s41524-023-01015-6
(17) Kim, S.; Hood, S. N.; Park, J.-S.; Whalley, L. D.; Walsh, A. Quick-Start Guide for First-Principles Modelling of Point Defects in Crystalline Materials. *J. Phys.: Energy* **2020**, *2* (3), 036001. https://doi.org/10.1088/2515-7655/aba081
(18) Squires, A. G.; Kavanagh, S. R.; Walsh, A.; Scanlon, D. O. Guidelines for Robust and Reproducible Point Defect Simulations in Crystals. *Nat. Rev. Mater.* **2026**. https://doi.org/10.1038/s41578-025-00879-y
(19) Dreyer, C. E.; Janotti, A.; Lyons, J. L.; Wickramaratne, D. Defects in Semiconductors. *J. Appl. Phys.* **2024**, *136* (19), 190401. https://doi.org/10.1063/5.0244142
(20) Mosquera-Lois, I.; Kavanagh, S. R.; Klarbring, J.; Tolborg, K.; Walsh, A. Imperfections Are Not 0 K: Free Energy of Point Defects in Crystals. *Chem. Soc. Rev.* **2023**, *52* (17), 5812–5826. https://doi.org/10.1039/D3CS00432E
(21) Rogal, J.; Divinski, S. V.; Finnis, M. W.; Glensk, A.; Neugebauer, J.; Perepezko, J. H.; Schuwalow, S.; Sluiter, M. H. F.; Sundman, B. Perspectives on Point Defect Thermodynamics. *phys. status solidi (b)* **2014**, *251* (1), 97–129. https://doi.org/10.1002/pssb.201350155
(22) Scheerer, O.; Juda, U.; Höhne, M. Self-Interstitial Shallow-Donor Complexes in Silicon: An Electron-Paramagnetic-Resonance Study. *Phys. Rev. B* **1998**, *57* (16), 9657–9662. https://doi.org/10.1103/PhysRevB.57.9657
(23) Tin, C.-C. Deep Level Transient Spectroscopy; Kaufmann, E. N., Ed.; Wiley, 2012; pp 1–14. https://doi.org/10.1002/0471266965.com036.pub2
(24) Binetti, S.; Le Donne, A.; Sassella, A. Photoluminescence and Infrared Spectroscopy for the Study of Defects in Silicon for Photovoltaic Applications. *Sol. Energy Mater. Sol. Cells* **2014**, *130*, 696–703. https://doi.org/10.1016/j.solmat.2014.02.004
(25) Mosquera-Lois, I.; Kavanagh, S. R.; Walsh, A.; Scanlon, D. O. Identifying the Ground State Structures of Point Defects in Solids. *npj Comput. Mater.* **2023**, *9* (1), 25. https://doi.org/10.1038/s41524-023-00973-1
(26) Huang, S.-D.; Shang, C.; Kang, P.-L.; Zhang, X.-J.; Liu, Z.-P. LASP: Fast Global Potential Energy Surface Exploration. *Wires Comput. Mol. Sci.* **2019**, *9* (6), e1415. https://doi.org/10.1002/wcms.1415
(27) Shang, C.; Liu, Z.-P. Stochastic Surface Walking Method for Structure Prediction and Pathway Searching. *J. Chem. Theory Comput.* **2013**, *9* (3), 1838–1845. https://doi.org/10.1021/ct301010b
(28) Perdew, J. P.; Burke, K.; Ernzerhof, M. Generalized Gradient Approximation Made Simple. *Phys. Rev. Lett.* **1996**, *77* (18), 3865–3868. https://doi.org/10.1103/PhysRevLett.77.3865
(29) Sun, J.; Ruzsinszky, A.; Perdew, J. P. Strongly Constrained and Appropriately Normed Semilocal Density Functional. *Phys. Rev. Lett.* **2015**, *115* (3), 036402. https://doi.org/10.1103/PhysRevLett.115.036402
(30) Grimme, S.; Antony, J.; Ehrlich, S.; Krieg, H. A Consistent and Accurate *Ab Initio* Parametrization of Density Functional Dispersion Correction (DFT-D) for the 94 Elements H-Pu. *The Journal of Chemical Physics* **2010**, *132* (15), 154104. https://doi.org/10.1063/1.3382344
(31) Coutinho, J.; Markevich, V. P.; Peaker, A. R. Characterisation of Negative- *U* Defects in Semiconductors. *J. Phys.: Condens. Matter* **2020**, *32* (32), 323001. https://doi.org/10.1088/1361-648X/ab8091
(32) Vaitkus, J. V.; Rumbauskas, V.; Mockevicius, G.; Zasinas, E.; Mekys, A. An Evidence of Strong

Electron–Phonon Interaction in the Neutron Irradiation Induced Defects in Silicon. *Nucl. Instrum. Methods Phys. Res. A: Accel. Spectrom. Detect. Assoc. Equip.* **2015**, *796*, 114–117. https://doi.org/10.1016/j.nima.2015.03.061
(33) Khan, F. S.; Allen, P. B. Deformation Potentials and Electron-Phonon Scattering: Two New Theorems. *Phys. Rev. B* **1984**, *29* (6), 3341–3349. https://doi.org/10.1103/PhysRevB.29.3341
(34) Zeng, Y.; Yang, D.; Ma, X.; Chen, J.; Que, D. Oxygen Precipitation in Heavily Phosphorus-Doped Silicon Wafer Annealed at High Temperatures. *Materials Science and Engineering: B* **2009**, *159–160*, 145–148. https://doi.org/10.1016/j.mseb.2008.12.045
(35) Ganose, A. M.; Scanlon, D. O.; Walsh, A.; Hoye, R. L. Z. The Defect Challenge of Wide-Bandgap Semiconductors for Photovoltaics and Beyond. *Nat Commun* **2022**, *13* (1), 4715. https://doi.org/10.1038/s41467-022-32131-4
(36) Lyons, J. L.; Wickramaratne, D.; Janotti, A. Dopants and Defects in Ultra-Wide Bandgap Semiconductors. *Current Opinion in Solid State and Materials Science* **2024**, *30*, 101148. https://doi.org/10.1016/j.cossms.2024.101148
(37) Kresse, G.; Hafner, J. *Ab Initio* Molecular Dynamics for Open-Shell Transition Metals. *Phys. Rev. B* **1993**, *48* (17), 13115–13118. https://doi.org/10.1103/PhysRevB.48.13115
(38) Kresse, G.; Furthmüller, J. Efficient Iterative Schemes for *Ab Initio* Total-Energy Calculations Using a Plane-Wave Basis Set. *Phys. Rev. B* **1996**, *54* (16), 11169–11186. https://doi.org/10.1103/PhysRevB.54.11169
(39) Grimme, S.; Ehrlich, S.; Goerigk, L. Effect of the Damping Function in Dispersion Corrected Density Functional Theory. *J Comput Chem* **2011**, *32* (7), 1456–1465. https://doi.org/10.1002/jcc.21759
(40) Grimme, S.; Antony, J.; Ehrlich, S.; Krieg, H. A Consistent and Accurate *Ab Initio* Parametrization of Density Functional Dispersion Correction (DFT-D) for the 94 Elements H-Pu. *The Journal of Chemical Physics* **2010**, *132* (15), 154104. https://doi.org/10.1063/1.3382344
(41) Blöchl, P. E. Projector Augmented-Wave Method. *Phys. Rev. B* **1994**, *50* (24), 17953–17979. https://doi.org/10.1103/PhysRevB.50.17953
(42) Van De Walle, C. G.; Neugebauer, J. First-Principles Calculations for Defects and Impurities: Applications to III-Nitrides. *J. Appl. Phys.* **2004**, *95* (8), 3851–3879. https://doi.org/10.1063/1.1682673
(43) Van De Walle, C. G.; Blöchl, P. E. First-Principles Calculations of Hyperfine Parameters. *Phys. Rev. B* **1993**, *47* (8), 4244–4255. https://doi.org/10.1103/PhysRevB.47.4244
(44) Kumagai, Y.; Oba, F. Electrostatics-Based Finite-Size Corrections for First-Principles Point Defect Calculations. *Phys. Rev. B* **2014**, *89* (19), 195205. https://doi.org/10.1103/PhysRevB.89.195205
(45) Thomas C. Allison. NIST-JANAF Thermochemical Tables - SRD 13, 2013. https://doi.org/10.18434/T42S31
(46) Jain, M.; Chelikowsky, J. R.; Louie, S. G. Reliability of Hybrid Functionals in Predicting Band Gaps. *Phys. Rev. Lett.* **2011**, *107* (21), 216806. https://doi.org/10.1103/PhysRevLett.107.216806

## Author contributions

All authors discussed the results and contributed to the manuscript.

## Acknowledgements

We acknowledge funding supports from the Strategic Priority Research Program of the Chinese Academy of Sciences (Grant No. XDB0670100), and the National Natural Science Foundation of China (Grant No. 22403104).

**Conflict of Interest**

The authors declare no conflict of interest.